# A NOVEL DEEP CLUSTERING FRAMEWORK FOR FINE-SCALE PARCELLATION OF AMYGDALA USING DMRI TRACTOGRAPHY

*Haolin He[1], Ce Zhu[1], Le Zhang[1], Yipeng Liu[1], Xiao Xu[1], Yuqian Chen[2], Leo Zekelman[2], Jarrett Rushmore[3], Yogesh Rathi[2], Nikos Makris[2], Lauren J. O'Donnell[2], Fan Zhang[1]*

[1] University of Electronic Science and Technology of China, Chengdu, China
[2] Harvard Medical School, Boston, USA
[3] Boston University, Boston, USA

## ABSTRACT

The amygdala plays a vital role in emotional processing and exhibits structural diversity that necessitates fine-scale parcellation for a comprehensive understanding of its anatomico-functional correlations. Diffusion MRI tractography is an advanced imaging technique that can estimate the brain's white matter structural connectivity to potentially reveal the topography of the amygdala for studying its subdivisions. In this work, we present a deep clustering pipeline to perform automated, fine-scale parcellation of the amygdala using diffusion MRI tractography. First, we incorporate a newly proposed deep learning approach to enable accurate segmentation of the amygdala directly on the dMRI data. Next, we design a novel streamline clustering-based structural connectivity feature for a robust representation of voxels within the amygdala. Finally, we improve the popular joint dimensionality reduction and *k*-means clustering approach to enable amygdala parcellation at a finer scale. With the proposed method, we obtain nine unique amygdala parcels. Experiments show that these parcels can be consistently identified across subjects and have good correspondence to the widely used coarse-scale amygdala parcellation.

## 1. INTRODUCTION

The amygdala is a major center for human emotional processing and is vital in linking emotions to many other brain functions including memories, learning, and senses [1]. Anatomically, the amygdala is a collection of nuclei located within the temporal lobe [2], [3], and each of its subdivisions exhibits unique functional roles [4]. Studies have shown that the distinct functional specificity of the amygdala's subdivisions plays an important role in the progression and diagnosis of brain diseases such as autism, Alzheimer's disease, and anxiety [5]. Therefore, meticulous parcellation of the amygdala is of great interest for understanding its neuroanatomical features and can potentially assist diagnosis and therapy of brain diseases.

MRI is a widely used imaging technique to study the amygdala and its subdivisions. The most commonly used is the structural MRI (sMRI), e.g., T1-weighted, from which segmentation[1] of the entire amygdala can be computed [6]. To investigate the detailed nuclear structure, studies utilize advanced high-resolution sMRI scans for a finer-scale parcellation of the amygdala [7], [8], e.g., nine amygdala nuclei were delineated using postmortem specimens at high resolution (100–150 μm) on a 7T scanner [7]. However, one shortcoming of sMRI is that it does not provide information about the inter-regional connectivity of the brain. Therefore, there has been work trying to subdivide the amygdala based on functional activation or connectivity derived from function MRI (fMRI) [9]–[11]. E.g., one recent fMRI study [10] subdivides the amygdala into two parcels and shows good correspondence with the widely used parcellation atlas that includes the laterobasal (LB), the centromedial (CM), and the superficial (SF) subdivisions [12]. Another category of work attempts to use diffusion MRI (dMRI) to parcellate the amygdala based on its white matter (WM) structural connectivity, which falls in the interest of our study. dMRI is an advanced MR technique that enables *in vivo* reconstruction of the brain's WM connections, via a computational process called tractography [13], [14]. Several studies have used dMRI tractography to extract WM connections related to the amygdala to perform structural connectivity-based parcellation [15]–[17].

There are still several challenges in effective amygdala parcellation using dMRI tractography. The first is how to accurately segment the amygdala in dMRI data. The commonly used approach is segmenting the amygdala on sMRI and then registering it to the dMRI space. However, the registration is challenging due to image distortions and low resolution of dMRI data, often resulting in segmentation errors between the amygdala and its neighboring regions [18]. The second challenge is how to come up with an informative connectivity feature representation that can discriminatively describe the amygdala's subdivisions. Existing work usually represents voxels using local fiber orientation distribution [15] or connectivity probability derived from probabilistic tractography [16], [17]. However, these methods do not consider the actual trajectory of the WM fibers that reflect the underlying WM anatomy. The third challenge is how to effectively identify the subregions of the amygdala based on its connectivity features. This has been done by using traditional clustering approaches, e.g., k-means [15], [16], to identify voxels with similar connectivity patterns, while recent advances in deep clustering have shown superior performance over traditional methods in dMRI-related parcellation tasks [19], [20].

---

[1] We use "segmentation" to refer to delineating the entire amygdala and "parcellation" to refer to identifying its subdivisions.

In this paper, we propose a novel deep-learning method for automated, fine-scale parcellation of the amygdala using dMRI tractography. Our method includes three major contributions. First, we segment the entire amygdala directly from dMRI data to avoid any potential segmentation errors caused by inter-modality registration between sMRI and dMRI data (Sec. 2.1). Second, we design a novel streamline clustering-based connectivity feature to enable robust representation of the voxels within the segmented amygdala (Sec. 2.2). Third, we propose a deep clustering network that extends the popular joint dimensionality reduction and *k*-means clustering approach [21] to groups voxels with similar connectivity for amygdala parcellation at a finer scale (Sec. 2.3). Experimental results show that our method enables consistent amygdala parcellation into nine parcels across multiple subjects and achieves good correspondence with a widely used coarse-scale amygdala atlas.

## 2. METHODS

### 2.1 Dataset, preprocessing, and amygdala segmentation

We utilize dMRI data from 100 young healthy adults (29.1±3.7 years; 54 females and 46 males) from the Human Connectome Project (HCP) [22]. The acquisition parameters are TE=89.5ms, TR=520ms, and voxel size=1.25×1.25×1.25mm$^3$, 18 baseline images, and 270 diffusion-weighted images distributed evenly at b=1000/2000/3000 s/mm$^3$. The provided dMRI data has been preprocessed as in [23].

Segmentation of the entire amygdala is performed using our recently proposed DDParcel method [18] that performs the FreeSurfer Desikan-Killiany (DK) parcellation [24] including the amygdala. Unlike existing methods that perform segmentation on T1-weighted data and then register to dMRI space, DDParcel enables amygdala segmentation directly from the dMRI data to avoid potential errors due to inter-modality registration. In this way, we can obtain an accurate amygdala segmentation to benefit the following connectivity-based parcellation into multiple subdivisions.

### 2.2 Voxel-wise connectivity feature representation

We compute a structural connectivity feature representation of each voxel within the segmented amygdala based on its WM connections. Unlike existing work that usually represents voxels using local fiber orientation distribution [15] or connectivity probability derived from probabilistic tractography [16], [17], we propose to use deterministic tractography data parcellated into streamline clusters for a robust connectivity representation.

#### 2.2.1. Amygdala tractography and streamline clustering

We first compute amygdala tractography to extract all WM connections related to the amygdala. To do so, we perform whole-brain tractography using a dual-tensor Unscented Kalman Filter (UKF) method [25] (provided in SlicerDMRI [26], [27]), followed by extracting all streamlines connecting between the amygdala and other brain regions. Next, we perform streamline clustering of the amygdala tractography using our well-established *whitematteranalysis (WMA)* pipeline [28]–[30] for simultaneous streamline clustering across all 100 subjects under study. As a result, the amygdala tractography of each subject is subdivided into $K$ streamline clusters. In our study, $K$ is set to 150 (parameter search from 20 and 200), which generates a good generalization across subjects while largely separating streamlines with different trajectories.

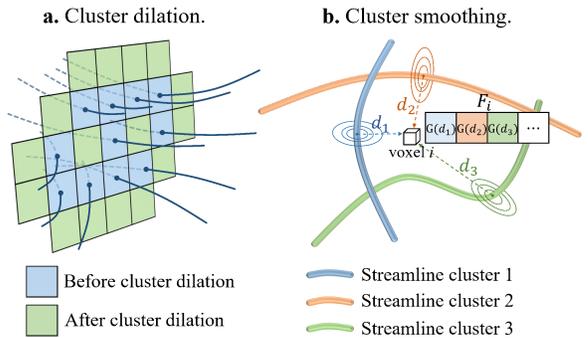

**Fig. 1.** Illustration of the cluster dilation and smoothing.

#### 2.2.2. Feature extraction via cluster dilation and smoothing

We then extract a feature vector for each voxel per subject based on the streamline clusters. Based on our previous work for dentate nucleus parcellation [20], for each voxel $i$, we construct a feature vector $F_i = (v_{ij} | j = 1, …, K)$, where $v_{ij}$ is set to 1 if streamline cluster $j$ intersects with $i$. However, this representation can be ineffective for brain structures like the amygdala which has highly complex WM connections. First, there are voxels within the amygdala not directly intersecting with any streamline clusters (referred to as *empty voxels*), resulting in $F_i$ being a zero vector. Second, this representation neglects the spatial relationship between the voxels and the clusters; that is to say, $v_{ij}$ is 0 no matter how far it is between $i$ and $j$. To resolve these, we propose two new additions for an improved feature representation.

First, we design a *cluster dilation* process, as illustrated in Fig. 1.a. For each empty voxel $i$, we set $v_{ij}$=1 if any adjacent voxel of *i* intersects with cluster *j*. This process largely reduces the sparsity effect of streamlines within a cluster, which results in voxels that are generally passed through the cluster but do not intersect with any individual streamlines. Second, we design a *cluster smoothing* process, as illustrated in Fig. 1.b. For each voxel *i* with $v_{ij} = 0$, we set $v_{ij} = G(d_{ij})$, where $d_{ij}$ is the smallest distance between *i* and all voxels intersecting with cluster *j*, and $G(d)$ is computed via a Gaussian smoothing with mean = 0 and σ = 1. After this, each element in $F_i$ includes the spatial distance between the voxels and the streamline clusters, while ensuring no empty voxels with a zero feature vector.

## 2.3 Fine-scale amygdala parcellation via deep clustering

We then parcellate the amygdala by clustering the voxels based on their connectivity features. To do this, we design a dense autoencoder network that extends the deep clustering method in [21] for an improved fine-scale parcellation. The overall network architecture is shown in Fig. 2, including three major parts: 1) *encoder* to extract a low-dimensional latent feature for each voxel based on its input connectivity feature, 2) *decoder* to reconstruct the original input from the low-dimensional feature, and 3) *k-means* to cluster voxels belonging to the segmented amygdala, thereby achieving amygdala parcellation into multiple parcels.

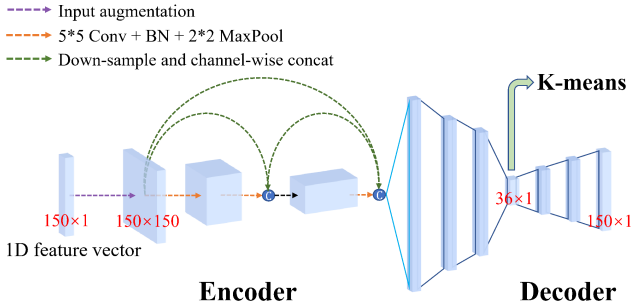

**Fig. 2.** Overview of the proposed network.

*2.3.1. Input augmentation and dense autoencoder*

Compared to the original network proposed in [21], we first include an input augmentation process to address the specialty of the input feature. Specifically, the original input is a 1D feature vector, where each element corresponds to one streamline cluster and its neighboring elements are two other random streamline clusters. A straightforward way is to apply a 1D convolution network, but it can only use the randomly assigned neighborhood information. Hence, similar to our previous work that builds a 2D representation from a 1D vector for tractography parcellation [31], we perform an input augmentation by shuffling the elements in the feature vector sequentially and concatenating them into a 2D input matrix. In this way, a 2D convolution network can be applied to use information from more streamline clusters of a target element. In our study, we cyclically shifted the feature vector *K* (the number of streamline clusters) times, generating a feature matrix with size 150×150. In addition, we improve the original simple convolution network with a DenseNet structure [32] in the encoder to better use the input feature, as illustrated with the green dashed arrows in Fig. 3.

*2.3.2. Adaptive k-means-friendly training mechanism*

One critical challenge for the deep clustering model is to ensure that the learned latent features from pre-trained autoencoders can be adapted according to the downstream clustering task. While the joint dimensionality reduction and *k*-means clustering approach in [21] has largely addressed this, we find it ineffective when a fine-scale clustering is performed, generating unbalanced clustering results where a subset of clusters is empty or with a small number of voxels. Therefore, we propose an improved adaptive training mechanism to avoid this issue, as follows. We first pre-train the autoencoder using a mean squared error (MSE) loss to learn a latent feature per voxel. Then, we simultaneously train the autoencoder and the *k*-means clustering based on the latent feature computed from the pre-training stage. This allows a fine-tuning of the autoencoder together with the clustering. During each training batch, we check if there are any clusters with only a small number of voxels (fewer than 1/80 of the batch size which is set to be 3096), and if so, replace these clusters' centroids with the mean of the others. Below is the loss function including one component for the autoencoder and one for cluster centroid assignment:

$$Loss_{train} = \mathrm{MSE}(d(e(x_i)), x_i) + \frac{\lambda}{2}\|e(x_i) - m_{k \leftarrow i}\|_2^2,$$

where $e(\cdot)$ is the encoder, $d(\cdot)$ is the decoder, $x_i$ is the input feature, $m_{k \leftarrow i}$ is the cluster centroid to which $x_i$ belongs, and λ= 3.3e-5 to balance the two components in the loss.

## 3. EXPERIMENTS AND RESULTS

### 3.1 Comparison with other parcellation methods

We compare the following methods: 1) the baseline *k-means* method that applies the traditional *k*-means clustering method directly on the input connectivity profile, 2) the *feat-orig* method that applies the proposed network on the original feature vector without the fiber dilation and smooth process, 3) the *net-orig* method that applies the original deep clustering network on our proposed feature, and 4) our method that applies the proposed network to the proposed feature. We use three evaluation metrics: 1) *Spatial Continuity (SC)* is the percentage of voxels in the maximum connected component per parcel, where a higher *SC* indicates a better spatial continuity, 2) *Parcel Size Coherence (PSC)* is the standard deviation of the sizes across the parcels per subject, where a low value indicates a highly uniformed parcel size, 3) *Dice Coefficient* (*Dice*) measures the spatial overlap of the corresponding parcels across subjects [33], where a high value indicates a high consistency. For each method, 80 subjects are used for training and 20 subjects are used for testing. For SC and PSC, we compute the mean value across all testing subjects. For Dice, we compute the mean score across all testing subject pairs for each parcel and also the average Dice across all parcels. We determined the number of parcels, $n$, by testing a range of values larger than 2 according to Dice. Dice decreases gradually as $n$ decreases but drops dramatically after $n = 9$. Therefore, we choose $n = 9$ as it gives the possibly finest parcellation scale. Table 1 gives the comparison results, where our method in general obtains the best performance across all compared methods.

Table 1. Quantitative comparison across different methods.

| | SC | PSC | Dice | | | | | | | | | |
|---|---|---|---|---|---|---|---|---|---|---|---|---|
| | | | Avg | $p_1$ | $p_2$ | $p_3$ | $p_4$ | $p_5$ | $p_6$ | $p_7$ | $p_8$ | $p_9$ |
| k-means | 0.82 | 0.19 | 0.27 | 0.19 | 0.16 | 0.29 | 0.28 | 0.24 | 0.16 | 0.26 | 0.21 | 0.56 |
| feat-orig | 0.82 | 0.21 | 0.21 | 0.11 | 0.17 | 0.21 | 0.23 | 0.06 | 0.05 | 0.27 | 0.59 | - |
| net-orig | 0.98 | 0.06 | 0.47 | 0.62 | 0.59 | 0.58 | 0.12 | 0.59 | 0.38 | 0.41 | 0.51 | 0.45 |
| ours | **0.99** | **0.03** | **0.51** | 0.59 | 0.57 | 0.62 | 0.56 | 0.49 | 0.19 | 0.49 | 0.55 | 0.55 |

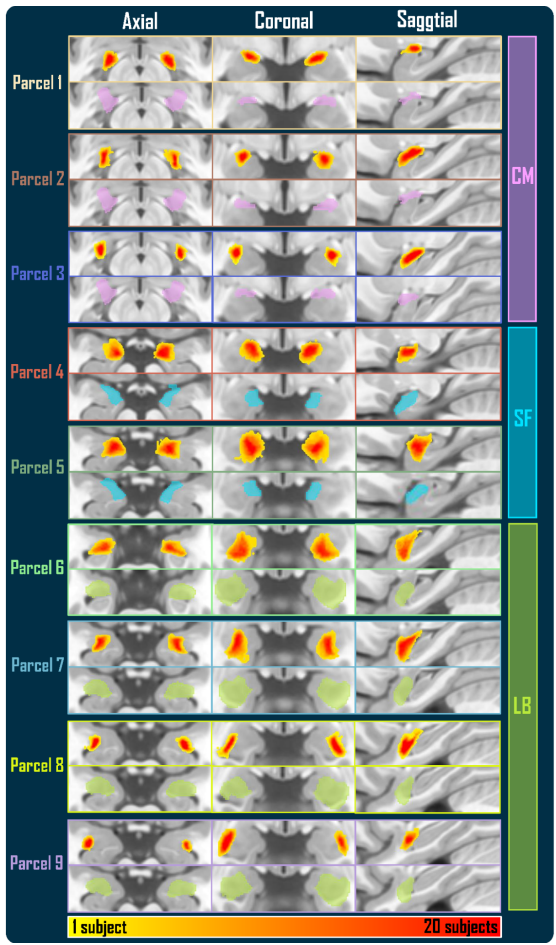

**Fig. 3.** Visualization of the group-wise parcellation result. For each parcel, the top row shows our result and the bottom row shows the corresponding SPM amygdala subdivision.

## 3.2 Comparison with existing parcellation atlas

To assess if our parcellation corresponds to the known anatomy of the amygdala, we compare our results with a widely used amygdala atlas that includes three major subdivisions, i.e., the laterobasal (LB), the centromedial (CM), and the superficial (SF) subdivisions, provided in the SPM Anatomy Toolbox [12]. Specifically, we register the dMRI data together with all parcels of each testing subject into the MNI space (i.e., the amygdala atlas space). Then, for each parcel $p$, we compute a heatmap where each voxel $v$ represents the number of subjects with $v$ parcellated to be $p$. Next, we binarize the heatmaps considering a voxel with a value over 10 belonging to the parcel on a group basis and compute the Dice overlap with each of the atlas subdivision.

Fig. 3 and Fig. 4 give a visualization of our group-wise and subject-specific parcellation of the amygdala, respectively, as well as the SPM amygdala atlas. In general, we have three parcels for CM, two for SF, and four for LB, each highly visually overlapping with the corresponding atlas parcel. Quantitatively, the average Dice between our parcels and the corresponding atlas parcels is 0.76, showing good correspondence between the two parcellation schemes.

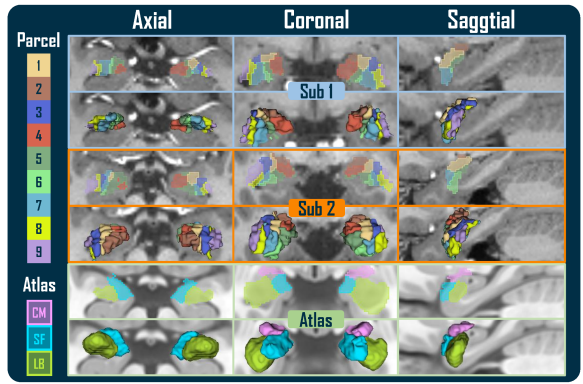

**Fig. 4.** Visualization of the parcellation results of two randomly selected subjects and the amygdala atlas. Top: 2D slice-wise visualization; bottom: 3D visualization.

## 4. DISCUSSION AND CONCLUSION

In this work, we propose a novel deep clustering pipeline to perform parcellation of the amygdala with dMRI tractography. We identify nine unique amygdala parcels that can be consistently identified across multiple subjects and have good correspondence to the known anatomical subdivisions of the amygdala. Further work can include an investigation of a finer-scale parcellation into more parcels and the application of our proposed method to other brain structures such as the thalamus and the striatum. Overall, we show the ability to use dMRI tractography to perform structural connectivity-based parcellation of the amygdala, which may provide insights into the detailed structure and function of the amygdala.

## 5. COMPLIANCE WITH ETHICAL STANDARDS

This study was conducted retrospectively using public HCP imaging data. No ethical approval was required.

## 6 ACKNOWLEDGEMENTS

This work is in part supported by the National Key R&D Program of China (No. 2023YFE0118600), the National Natural Science Foundation of China (No. 62371107) and the National Institutes of Health (R01MH125860, R01MH119222, R01MH132610, R01NS125781).